# Assessing the Socio-economic Impacts of Secure Texting and Anti-Jamming Technologies in Non-Cooperative Networks

OSORO B OGUTU[1], EDWARD J OUGHTON[1] (Member, IEEE), KAI ZENG[1] (Member, IEEE) AND BRIAN L. MARK[1](Senior Member, IEEE)
[1] George Mason University, Fairfax, VA 22030, USA, e-mail: (bosoro@gmu.edu)
[2] George Mason University, Fairfax, VA 22030, USA, e-mail: (eoughton@gmu.edu)

Corresponding author: Osoro B. Ogutu (e-mail: bosoro@gmu.edu).

This work was supported by the Geography & Geoinformation Science Department at George Mason University.

**ABSTRACT** Operating securely over 5G (and legacy) infrastructure is a challenge. In non-cooperative networks, malicious actors may try to decipher, block encrypted messages, or specifically jam wireless radio systems. Such activities can disrupt operations, from causing minor inconvenience, through to fully paralyzing the functionality of critical infrastructure. While technological mitigation measures do exist, there are very few methods capable of assessing the socio-economic impacts from different mitigation strategies. This leads to a lack of robust evidence to inform cost-benefit analysis, and thus support decision makers in industry and government. Consequently, this paper presents two open-source simulation models for assessing the socio-economic impacts of operating in untrusted non-cooperative networks. The first focuses on using multiple non-cooperative networks to transmit a message. The second model simulates a case where a message is converted into alternative plain language to avoid detection, separated into different portions and then transmitted over multiple non-cooperative networks. A probabilistic simulation of the two models is performed for a 15 km × 15 km spatial grid with 5 untrusted non-cooperative networks and intercepting agents. The results are used to estimate economic losses for private, commercial, government and military sectors. The highest probabilistic total losses for military applications include US\$300 $\pm$ 25, US\$150 $\pm$ 15, and US\$75 $\pm$ 10, incurred for a 1, 3 and 5 site multi-transmission approach, respectively, for non-cooperative networks when considering 1,000 texts being sent. These results form a framework for deterministic socio-economic impact analysis of using non-cooperative networks and secure texting as protection against radio network attacks. The simulation data and the open-source codebase is provided for reproducibility.

## I. INTRODUCTION

Ever since its invention and subsequent application, wireless radio communication has been at the center of world's technology. The ability to communicate wirelessly with distant parts of the world and beyond has made radio communication a fundamental technology supporting a wide range of use cases. For instance, 2.2 trillion text messages were sent in United States in 2020 alone [1]. However, the ability to connect two or more places remotely is a ready testbed for adversaries to penetrate new networks with the aim of stealing information, spying, sabotage, or attack. Consequently, radio systems can be a single point of failure for information interception, information transfer prevention or causing the disruption to critical infrastructure.

Indeed, society's reliance on radio technology makes this group of technologies an easy target for malicious actors to cause disruption and other negative outcomes. Jamming is a common phenomenon in compromising the integrity and usability of such networks [2]. Different forms of jamming such as suppression and deception have been widely used to counter radio-reliant systems such as radar [3] and mobile edge computing devices [4] creating a whole domain of electronic warfare [5]. Indeed, jamming as a warfare tool has been in existence for almost a century, ever since the second world war, where it took place between attacking bomber formations and air defense systems [6]. Therefore, creating resilient radio networks has been an active research area for many decades.

As a consequence, research effort is being directed towards developing systems that do not alert malicious actors to signals of interest. The rationale for such systems is the need for governments, businesses, law enforcement and the military to have reliable, secure and covert communication systems that function in non-cooperative environments. Consequently, use of covert texting options have been explored in the literature [7], with the aim of guaranteeing that an intended message is only received by the target receiver.

The advent of big data and Internet of Things (IoT) have particularly accelerated the research on secure texting. For instance, [8] and [9] have proposed a schema for guaranteeing end-to-end message encryption in an IoT network. Additionally, [10] have proposed cryptosystems for encrypting sensitive healthcare data. Each of the proposed solutions seek to guarantee security and the integrity of the message to the target receiver. The implementation becomes complex when the solutions have to be deployed over public networks. However, a review by [11] indicate significant steps towards the deployment of the techniques on public networks. In a related study, [12] developed an algorithm that only allows intended users in an open network to read the message. A more empirical work involving the testing of an encryption on a public network has been conducted by [13].

Still, limited research exists on quantifying the effectiveness of secure texting over public networks. The advent of 5G and higher technologies provides an opportunity for exploring secure texting over multiple untrusted public networks. Through beamforming, scheduling and dynamic resource allocation, 5G devices can transmit and receive messages from multiple location and sources as enumerated in the literature [14], [15].

However, it is not only technical issues that need addressing. Issues pertaining to policy, economics and intellectual property are also highly important, and require analytics on radio security techniques to help service providers and governments to prioritize resources accordingly. Moreover, such analytics enables operators, regulatory bodies and vendors to implement and distribute the new enhanced security features based on reasonable cost-benefit analysis. Only limited cost benefit analysis exists in the literature for radio security techniques.

Therefore, there is need to quantify the success rate of applying spatial anti-jamming techniques and secure texting over multiple public untrusted networks before conducting their cost benefit analysis. In this paper, two models are developed that (i) simulate transmission of a single message over public multiple networks and (ii) simulate secure texting over non-cooperative networks and associated socio-economic costs.

The first model denotes a radio environment with multiple pairs (uplink/downlink) of beams between the transmitter and the user. The pair with the highest signal to noise plus interference ratio (SNIR) and orthogonal to the jammer's beam provides the best solution. The second model represents a case where a sensitive message is made to look like ordinary text through application of natural language processing (NLP) before separation and sending to multiple non-cooperative public networks.

Indeed, the evaluation of such preventive techniques should not be limited purely to the technical engineering results. Going one step further, the findings can be translated to potential socio-economic impacts, to aid technology evaluation and selection. Consequently, this work simulates the effectiveness of the proposed solutions and the accompanying socio-economic costs. Thus, the key research questions of this paper include:

1. What is the impact of anti-jamming techniques on texting in non-cooperative networks?
2. What is the impact of secure texting and multi-network sending in non-cooperative networks?
3. What are the socio-economic impacts of these techniques?

Having posed the research questions, a literature review of attacks on wireless radio systems is conducted in Section II, followed by the methodology in Section III, before results are presented in Section IV. The discussion of the results is made in Section V and conclusion in Section VI.

## II. LITERATURE REVIEW

As we increasingly rely on wireless radio systems for communication, the potential negative impacts associated with disruption to these technologies grow. To secure these systems we must begin by designing, operating, and modeling all possible attack scenarios, to enable a more proactive and strategic stance to be taken against malicious attacks [16]. Currently, protection strategies range from simple mathematical models which simulate the action of the malicious actors to complex techniques based on machine learning (ML) which attempt to actively identify irregular activity associated with malicious attackers [4].

TABLE 1
RADIO NETWORK ATTACK LITERATURE REVIEW

| Author | Year | Application | Contribution | Approach |
|---|---|---|---|---|
| Alhayani et al. [17] | 2021 | General | Cybersecurity Awareness | Review |
| Pycroft and Aziz [18] | 2018 | Healthcare | Solutions to Cyber Threats | Review |
| Sethuraman et al. [19] | 2019 | Healthcare | Vulnerability Testing | Empirical |
| Mualla et al. [20] | 2019 | UAVs | Future Research Directions | Systematic Literature Review |
| Oughton et al. [21] | 2017 | Energy | Risk Analysis | Simulation |
| Chhaya et al. [22] | 2017 | Energy | Intrusion Detection System | Assessment |
| Al-Dabbagh et al. [23] | 2018 | Energy | Control System Topology Design | Empirical |
| Buinevich and Vladyko [24] | 2019 | Transport | Analytical Overview | Case Study |
| Ghelani et al. [25] | 2022 | Banking | Cloud Data Security | Machine Learning (ML) |
| Melgarejo et al. [26] | 2022 | Communication | Interference on IoT Network | Simulation and AI |
| Lu et al. [27] | 2019 | Energy | Wireless Control System Solution | ML (Reinforcement Learning) |
| Kavousi-Fard et al. [28] | 2021 | Energy | Attack Detection | ML (Anomaly Detection) |
| Ismail et al. [29] | 2021 | Energy | ML in Cyber-attack detection | Comparative Study |
| Alhayani et al. [17] | 2021 | General | Cybersecurity Awareness | Review |

## A. ATTACKS ON WIRELESS RADIO SYSTEMS

The key to radio network attacks is anonymity. Radio systems provide a ready testbed for malicious agents to intrude and infect systems, without being physically present near to a radio asset (indeed, malicious agents could be on another continent). The remoteness of the activity allows for execution of phishing, interference, backdoors, and denial of service among other types of attacks [17]. Due to the complex nature of averting wireless attacks, research has focused on proactive mitigation strategies, for example, by focusing on modeling attack scenarios, and in some cases deploying them to test the impact on real systems [30]. Pycroft and Aziz have demonstrated the threat and impact of attacks on wireless implantable medical devices [18]. Their work not only highlights the vulnerability of wireless systems to attacks but also underscore the importance of protecting such systems. A more empirical, yet similar study, was conducted to model attacks on healthcare devices [19]. Rather than just modeling the attacks, the authors deployed a wireless attack from an uncrewed aerial vehicle (UAV) to a simulated hospital environment to quantify the effects. Indeed, involvement of UAVs in radio attack is not new as [20] has extensively reviewed their role in civilian radio applications.

Similarly, wireless sensor networks (WSNs) are radio-based systems which have experienced multiple different radio attacks too. For example, these technologies have attracted the attention of developers and researchers because a major disruption of such systems can lead to the loss of billions of dollars to businesses, households, governments and the wider economy. Chhaya et al. have proposed a topology control method as a solution for protecting such networks [22]. Their work has been further extended by [23] who developed a system of nodes that detected attacks at different scenarios. Related research in the domain of intelligent transport systems has been conducted to propose solutions against radio attacks that can lead to failure of automotive engineering and even death [24]. The general approach in these studies reveals that researchers are going beyond just studying attacks but also modeling, simulating and even empirically testing them on controlled systems to understand the impact.

Increasingly researchers are now harnessing the power of ML techniques to propose solutions against radio attacks and strategies [31]. For instance, [26] has applied reinforcement learning algorithms models to reduce unintentional inter-cell interference in an agricultural application IoT system. Moreover, [27] combined reinforcement learning algorithm with agent-based models to suggest the best traffic control solution in a wireless sensor IoT network. IoT networks benefit extensively from ML techniques to identify and prevent radio attacks due to the nature of sensor information they provide. The ML algorithms can detect anomalies as well as differentiate the severity of the attacks as demonstrated by [28]. In addition to individual techniques, [29] has conducted a comparative study on the best algorithms for detecting and mitigating attacks. Such studies reveal the power of new computational and legacy techniques in radio security research.

## B. CHALLENGES AND SOLUTIONS TO RADIO NETWORK ATTACKS

The security of radio networks is a growing research theme due to the development of new radio systems, such as 5G. Indeed, such developments require holistic assessment of the technology architecture, benefits, challenges and recommendations [32], [33]. Additional technologies such as Internet of Nano-Things and ML have been studied to boost the security and management of 5G radio networks [34]. For instance, [35] has demonstrated the application of convolution neural networks in building covert radio communication systems. Contrasting results by [36] indicated that ML methods can equally be used to compromise radio networks, specifically 5G. Irrespective of the approach, the focus has been on protecting the integrity of the network leading to enhanced encryption technologies that focus on hiding the users identity or blocking authorized access [37].

Randomized dynamic resource allocation (DRA) capability of 5G is now being fronted as a possible solution for guaranteeing secure communications especially in jamming environments due to the shortcomings of traditional strategies [38]. Jamming can be a threat to public safety especially in new systems such as IoT that have been integrated into utility services including water, electricity and transportation [39]. Despite the continual threats, the progress in designing resilient radio network through DRA has been slower than expected [40]. The slow progress has led to unconventional strategies such as integrating UAVs and implementing software defined networks to lower detection probability and accuracy [41], [42]. Authors [43] have for instance explored the idea of using a jammer's signals to help in spectrum coordination. In a related work, [30] simulated a radio environment where UAVs are used to avoid unauthorized access of information transmitted to the receiver. These studies underscore the importance of DRA and equally the limitations as a technique for protecting the integrity of radio networks.

Other than DRA, using multiple transmission and reception point (mTRP) to secure radio networks has become interfering beam mTRP is an effective strategy in a jamming environment.

### C. SOCIO-ECONOMIC IMPACT ANALYSIS OF RADIO SYSTEMS

Today, most of the government, commercial, social, cultural and economic interactions and activities between individuals, institutions, businesses and countries are conducted in cyberspace [48]. As a result, the security of radio networks is increasingly essential. Relationally, efforts are also directed at modeling, simulating, and quantifying the economic costs associated with radio network attacks. Research has shown that attacks not only affect targeted entities but also cascades to others and the general economy [49]. For instance, [50] developed a model to estimate the potential cascading economic damages of an attack and applied it to the insurance market. Despite these attempts to quantify the economic costs, most of the studies focus on critical infrastructure systems.

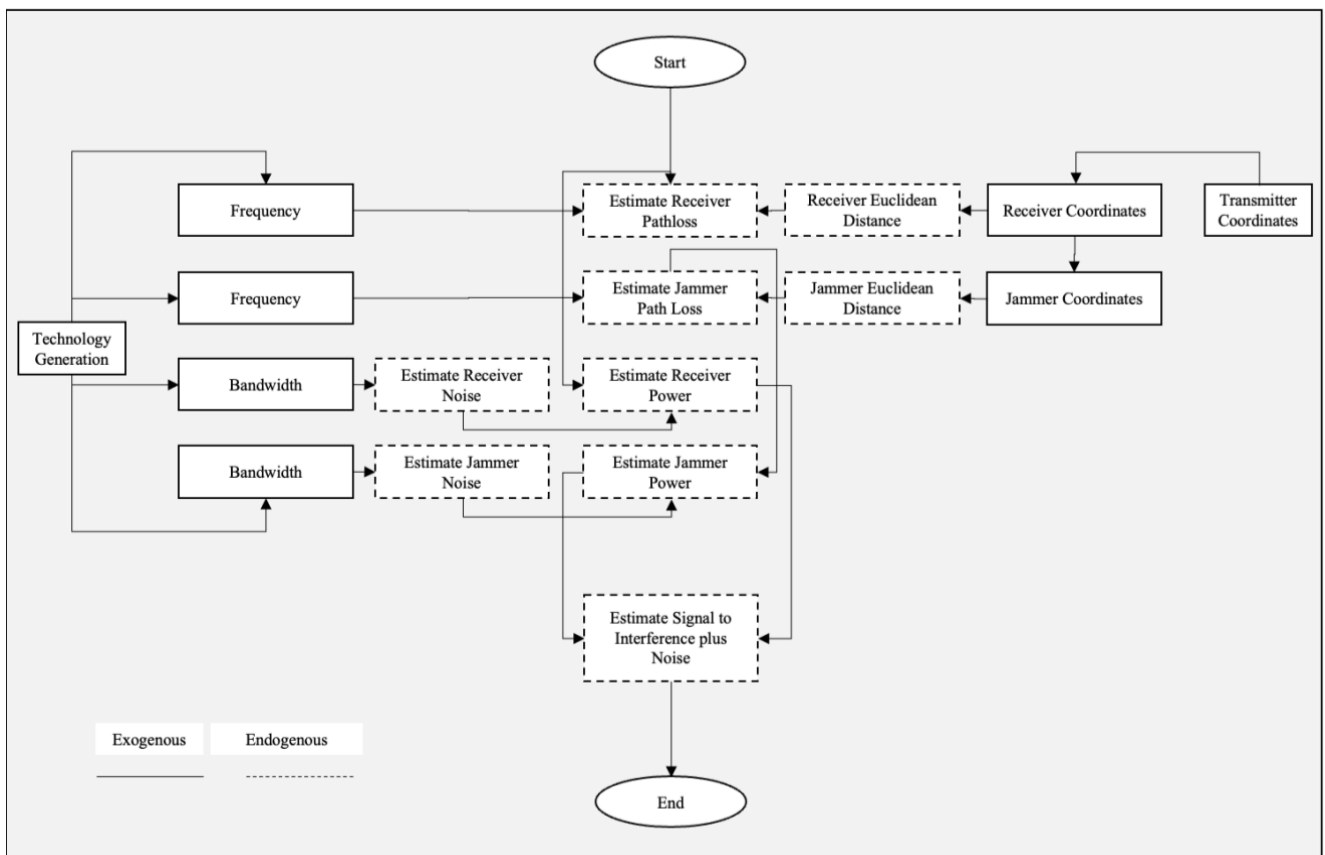

**FIGURE 1. Spatial jamming simulation framework**

popular [44], [45]. However, much of the mTRP work has focused on using the technique as a means of improving the quality of service (QoS) in a radio interference environment characterized by non-cooperative transmitters [15], [46]. The method limits the mutual interference between coexisting transmitters [47]. Therefore, using signals orthogonal to the

The goal of such research is to estimate the losses in situations of disruptions due to cyber-attacks. The quantification of losses due to false data injections on power systems has particularly gained popularity in this domain [51], [52]. For instance, [53] simulated a set of resilient strategies to guarantee that a system remains stable and

VOLUME XX, 2017

economically optimal when faced with data injected attacks.

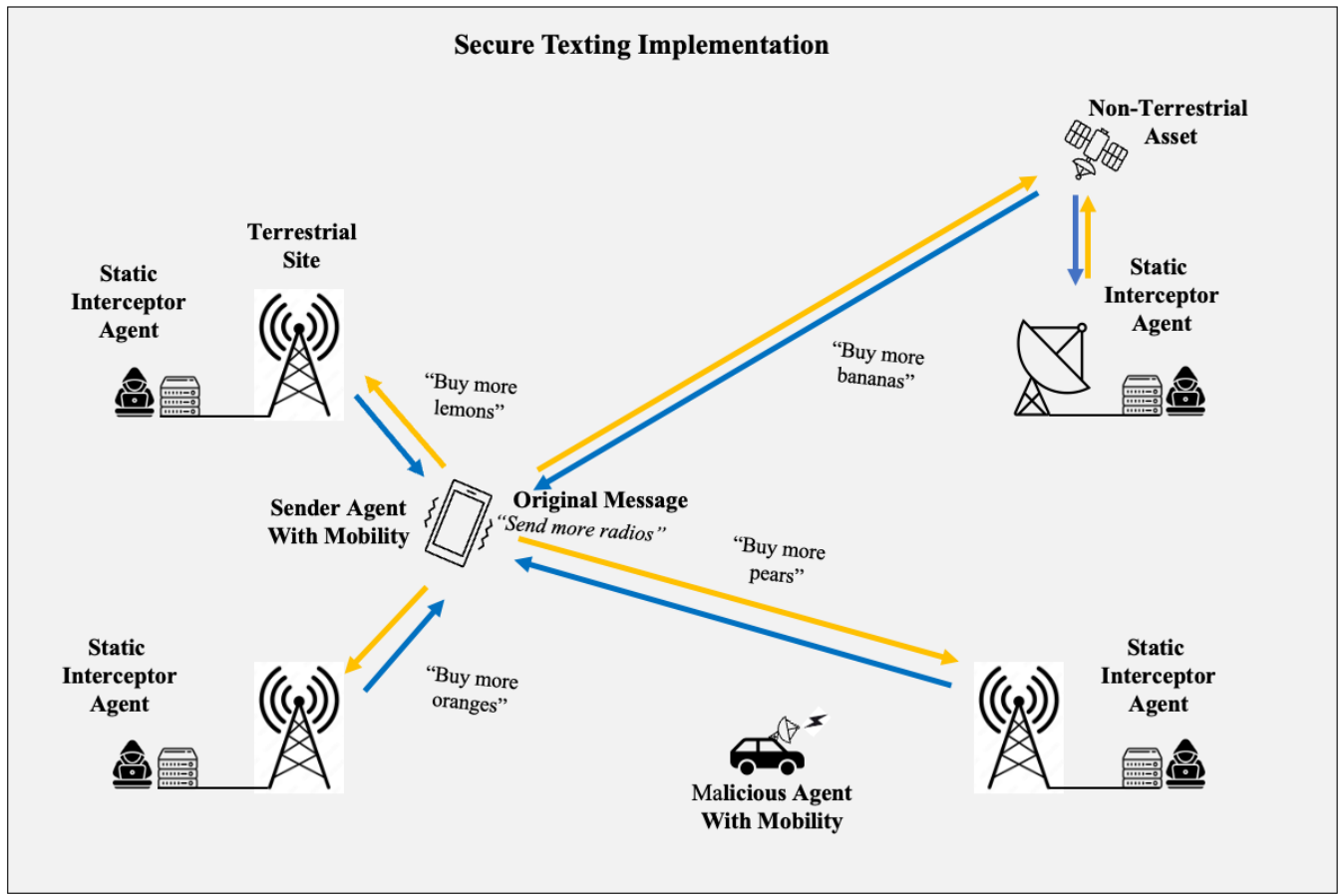

**FIGURE 2.** Covert secure texting framework implementation example.

In a related work, [54] assessed the economic impacts of cyber-attacks on renewable power systems and concluded that it increases total operating costs due to the need for commissioning expensive units. However, none of these studies focus exclusively on radio networks and losses associated with deploying different mitigation techniques. The attention is mostly on direct attacks such as false data injection or breaches on systems containing sensitive client data [55]. Even in the case of power grid systems, [56] posits that there is no general model for analyzing the cost of cascading failures due to a cyber-attack.

Limited open-source generalizable models exist that estimate the losses due to radio attacks alongside proposed mitigation strategies. Therefore, this paper seeks to simulate the application of covert texting and multi-transmission of messages in unsecure non-cooperative network as well as estimating the socio-economic costs. The simulation models are defined in the next section.

### III. METHOD
The methodology is broken down into two parts. The first part focuses on the case of a jammer blocking messages. The second part simulates secure texting by applying a combination of NLP and encryption techniques to hide messages in plain sight. In this approach, the original message is converted into everyday words to avoid being flagged. They are then separated into different sentences through NLP recommendation, grammatically corrected and transmitted into different networks. The reverse process is performed at the receiver's end.

### A. ANTI-JAMMING SIMULATION MODEL
Jamming involves transmitting similar (frequency and power) radio signals to the receiver thus saturating it with noise or false information. Generally, the closer the jammer to the target receiver, the higher the interference and hence loss of meaningful signal. In this paper, jamming of the network is treated as a spatial problem.

Consider a radio access network's (RAN) transmitter station $T_x$ located at $T_{(x,y)}$ serving a given cell. A user equipment (UE) $R_x$, located at $R_{(x,y)}$ connects to $T_x$ to relay the intended message. An interfering agent $I_x$ at position $I_{(x,y)}$ generates a replicate signal to block $R_x$ from accessing $T_x$.

On a business-as-usual (BAU) scenario where there is no interference, the signal to noise ratio (SNR) at $R_{(x,y)}$ is quantified by equation (1).

$$SNR(dB) = \frac{S_P}{N_P} \qquad (1)$$

Where $S_P$ is the transmitter signal power of $T_x$ and $N_P$ the noise power at $R_x$. The level of useful signal at $R_{(x,y)}$ is a

Where $f_{(Hz)}$ is the transmission frequency, $c_{(ms^{-1})}$ the speed summation of any unwanted interference power, $I_P$. Therefore, the resulting signal to interference plus noise ratio (SINR) is given by equation (2).

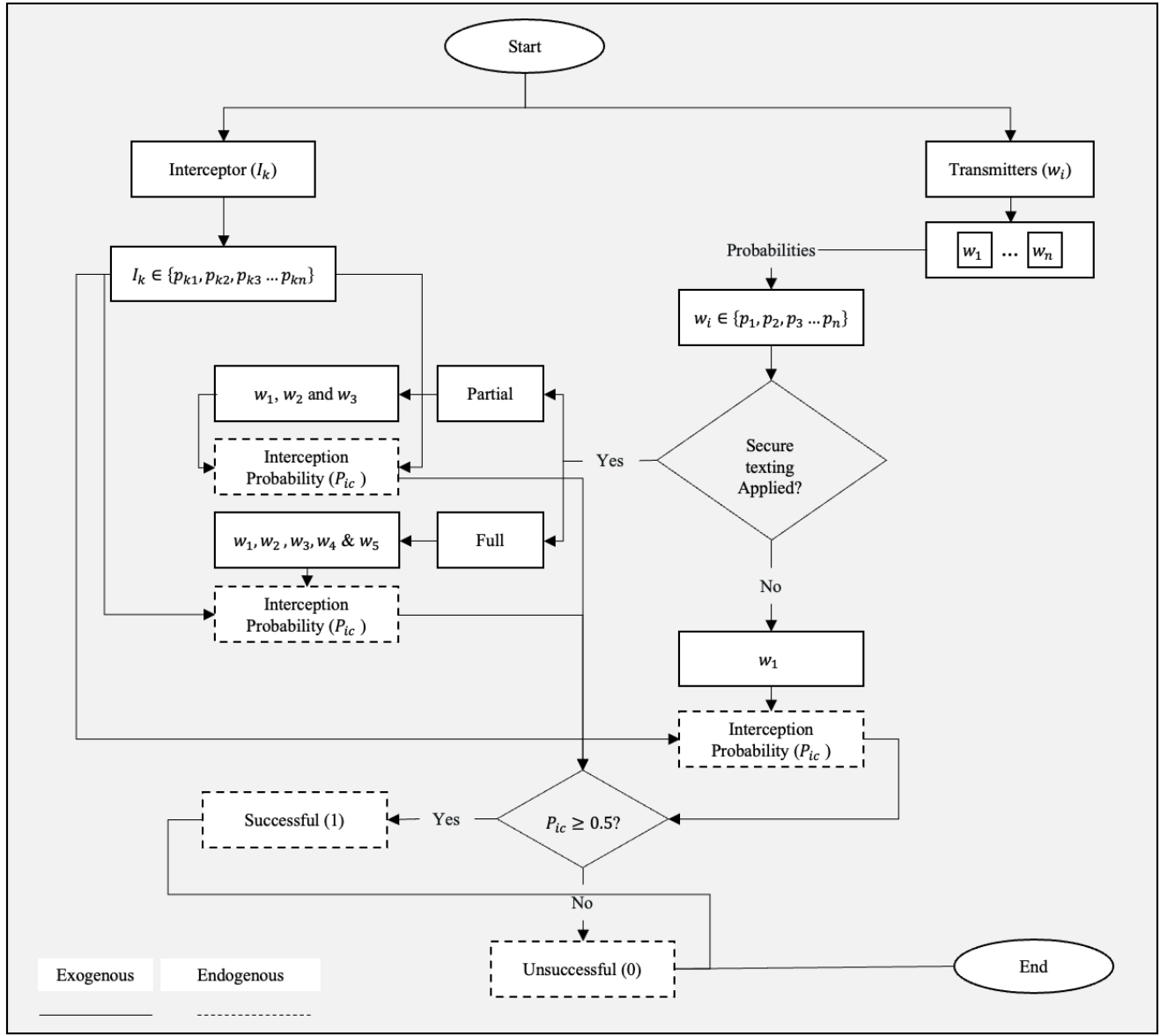

**FIGURE 3.** Covert secure texting implementation box diagram.

$$SINR(dB) = \frac{S_P}{I_P + N_P} \quad (2)$$

Both $I_P$ and $S_P$ are directly proportional to the Euclidean distance between $T_{(x,y)}$, $R_{(x,y)}$ and $I_{(x,y)}$ due to the path loss, $P_L$. The path loss can be quantified as in equation (3).

$$PL(dB) = 10 \log_{10}\left(\left(\frac{4\pi d_{(m)} f_{(Hz)}}{c_{(ms^{-1})}}\right)^2\right) \quad (3)$$

of light and $d_{(m)}$ the Euclidean distance between the transmitter and the receiver. The $d$ between $T_{(x,y)}$ and $R_{(x,y)}$ is different from that of $T_{(x,y)}$ and $R_{(x,y)}$ and is calculated using equations (4) and (5).

$$d_{TX} = \sqrt{\left[(R_x - T_x)^2 + (R_y - T_y)^2\right]} \quad (4)$$

$$d_{IX} = \sqrt{\left[(R_x - I_x)^2 + (R_y - I_y)^2\right]} \quad (5)$$

$d_{TX}$ represents the distance between the transmitter and receiver and it is responsible for the path loss that reduces the value of $S_P$. $d_{IX}$ accounts for the distance between the interfering agent and receiver and directly contributes to $I_P$. The shorter the distance, the higher the $I_P$ and subsequently the lower the $SINR$, potentially leading to jamming impacts. Taking into account the spatial changes for a given cell grid, $d_i \times d_j$, the jamming process can be treated as a spatial problem where the variation in receiver and interference distance leads to a series $SINR$ results.

For different sets of $T_{(x,y)}$, $R_{(x,y)}$ and $I_{(x,y)}$ the corresponding $S_P$, $I_P$ and $SINR$ can be computed to establish the best transmitter to connect to at a particular time instance. The simulation can be undertaken for different cellular technology generations (2G, 3G, 4G, and 5G) owing to the different operating frequencies and channel bandwidths. The spatial simulation model is illustrated graphically in Fig. 1.

### B. SECURE TEXTING STRATEGY
This section simulates the application of a secure texting in a jamming environment.

The secure texting in untrusted wireless network, $w_i$ with probabilities, $p$ of successfully sending messages is modeled. This can be presented with equation (6).

$$w_i = \{p_1, p_2, p_3 \dots p_n\} \quad (6)$$

For simplicity, the $p$'s are arbitrarily selected and every time a sender connects to the site with the best $SINR$, the success of the message is given a random probability from the $w_i$ set.

In the case of secure texting implementation, a set of untrusted networks, $w_u$ with the best $SINR$ exists.

$$w_u = \{w_1, w_2, w_3 \dots w_n\} \quad (7)$$

The interceptor agent, $I_k$ can read/block the messages with a random probability value, equation (8).

$$I_k = \{p_{k1}, p_{k2}, p_{k3} \dots p_{kn}\} \quad (8)$$

Therefore, the probability of intercepting, $P_{ic}$ the message is quantified using random choices from $w_i$ and $I_k$ as shown in equation (9).

$$P_{ic} = w_i - I_k \quad (9)$$

The BAU scenario occurs when the secure texting is not implemented and the message is transmitted on a single untrusted wireless network from the set, $w_u$. The $P_{ic}$ is a function of only a single site as shown in equation (9).

In the case of secure texting implementation, the $w_i$ can be from any network from the $w_u$ set. Hence $P_{ic}$, is calculated by equation (10).

$$P_{ic} = ([p(w_1) \dots p(w_n)] - I_k) \quad (10)$$

The message delivery probability, $M(P_{ic})$ is quantified by a probability threshold as set in equation (11).

$$M(P_{ic}) = \begin{cases} 1, & P_{ic} \geq 0.5 \\ 0, & P_{ic} < 0.5 \end{cases} \quad (11)$$

Where 1 and 0 represents successful and unsuccessful messages respectively.

### C. SOCIO-ECONOMIC SIMULATION
In this section, a simulation model for quantifying the cost for all intercepted and blocked messages is presented. The model is based on the approach used in power engineering economics quantified by value of lost load (VoLL). The VoLL represents the amount electricity users are willing to pay during disruption of service [57]. The VoLL is defined by equation (12).

$$Loss(\$/kW) = f(d_r, S_s, t_d, n) \quad (12)$$

Where $d_r$ is the duration, $S_s$ season, time of the day $t_d$ and $n$ notice. However, for the case of this paper we modify the VoLL as shown in equation (13).

$$Loss\ (\$/msg) = \frac{\alpha \cdot \beta \cdot C_{msg}}{M(P_{ic})} \quad (13)$$

$M(P_{ic})$ is defined from the secure texting model and $\alpha$ the and $\beta$ are coefficients representing the multi-transmission modes and application sector for which network is used for (personal, commercial, government or military) respectively. The values depend on usage in each application sector. For instance, the coefficient value for government should be high compared to personal use. The $C_{msg}$ is the cost corresponding to the probability of interception or blocking and generally follows the Poisson distribution. The distribution is based on empirically determined values from VoLL surveys, except that these values are related to power outage hours instead of probability [58], [59]. The calculation of $Loss\ (\$/msg)$ enables the quantification of the total costs by multiplying it with all the messages intercepted or blocked.

### D. APPLICATION
This section entails the simulation of the jamming and secure texting application. A hypothetical environment is simulated where users are trying to send a message through an untrusted wireless network. The simulation entails jamming of the network as well as implementation of secure texting.

A $15\ km \times 15\ km$ grid (assumed to be enclosing an hexagon) is selected as cellular signals tend to degrade rapidly beyond 10 kilometers distances. The grid consists of 5 base stations ($T_x$) each with given $(x, y)$ coordinates. The grid also has 5 static interceptor agents ($I_x$) each at $(x, y)$ location. The coordinate location of the interceptor and transmitter RAN base stations are presented in Table 3.

TABLE 3
TRANSMITTER AND INTERCEPTOR COORDINATES

| ID | X - Coordinates | Y - Coordinates | Agent |
|----|-----------------|-----------------|-------------|
| A  | 4               | 4               | Transmitter |
| B  | 4               | 4               | Transmitter |
| C  | 8               | 8               | Transmitter |
| D  | 14              | 14              | Transmitter |
| E  | 14              | 14              | Transmitter |
| A  | 4.1             | 14.1            | Interceptor |
| B  | 4.1             | 4.1             | Interceptor |
| C  | 8.1             | 8.1             | Interceptor |
| D  | 14.1            | 4.1             | Interceptor |
| E  | 14.1            | 14.1            | Interceptor |

The user can move anywhere in the grid to relay a mission critical message. Therefore, a simulation is performed for all integer position of the UE within a grid to determine, both the interferer's and receiver's distance, path loss and power to establish the best SINR. The secure texting approach is then simulated in a jamming environment. Five unsecure wireless networks ($w_1, w_2, w_3, w_4$, and $w_5$) are used with a single static interceptor, $I_1 = \{0.1, 0.2, 0.3, 0.4\}$. The arbitrary values chosen for the unsecure wireless networks are as follows.

the quantification of the loss by adopting the Poisson distributed cost values as shown in Table 4.

TABLE 4
COST LOOKUP TABLE

| Probability | Cost (US$) |
|---|---|
| <=0.1 | 10 |
| 0.2 | 7.2 |
| 0.3 | 5.1 |
| 0.4 | 2.8 |
| 0.5 | 1.1 |
| 0.6 | 0.1 |
| 0.7 | 0.1 |
| 0.8 | 0.1 |
| >=0.9 | 0.1 |

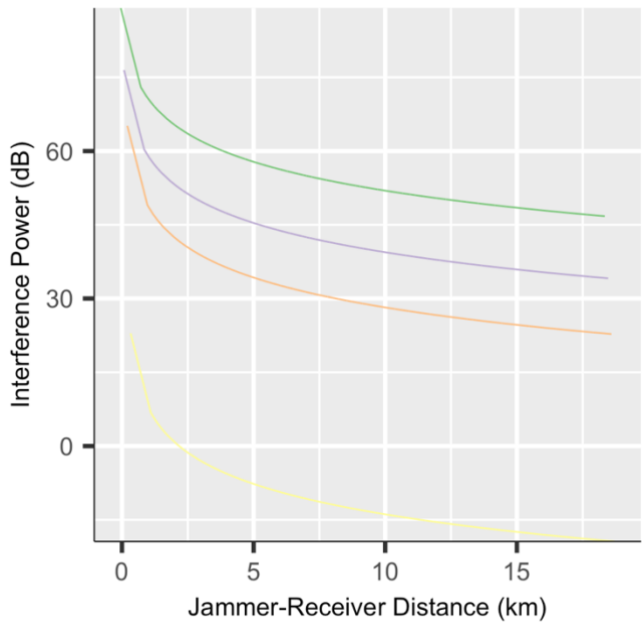
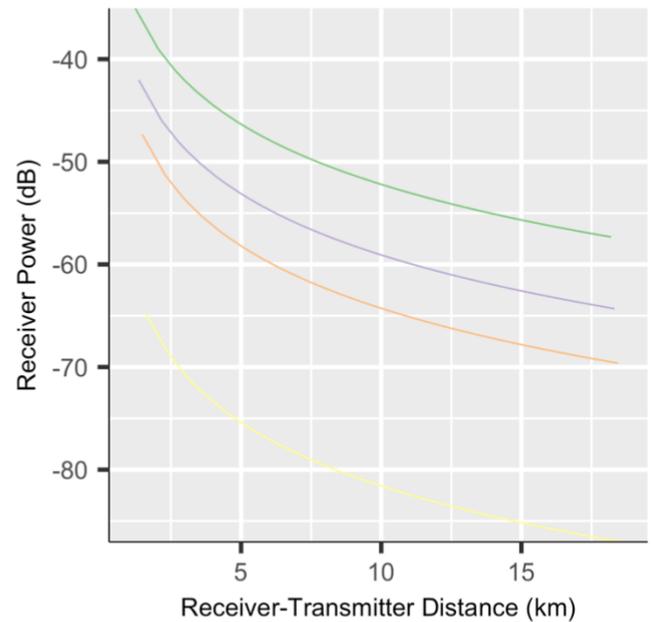
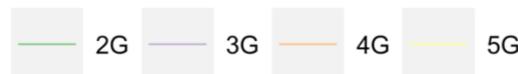

**FIGURE 4.** Radio signal results from the spatial modeling in a 15 km ×15 km grid with 5 transmitters and interceptors for different cellular generations.

$w_1 = \{0.1, \quad 0.2, \quad 0.3, \quad 0.4, \quad 0.45\}$
$w_2 = \{0.1, \quad 0.2, \quad 0.3, \quad 0.45, \quad 0.45\}$
$w_3 = \{0.2, \quad 0.2, \quad 0.3, \quad 0.45, \quad 0.45\}$
$w_4 = \{0.2, \quad 0.2, \quad 0.3, \quad 0.45, \quad 0.45\}$
$w_5 = \{0.2, \quad 0.2, \quad 0.3, \quad 0.45, \quad 0.45\}$

For the BAU scenario, only $w_1$ is used while $w_1$, $w_2$ and $w_3$ are used for the partial multi-transmission (3 sites). All unsecure networks ($w_1, w_2, w_3, w_4$ and $w_5$) are used for the case of full multi-transmission (5 sites). The results are used in

The multi-transmission modes are represented by coefficients 1, 2, and 3 for full, partial and baseline implementation respectively. The losses are then applied for selected application areas including private, commercial, government and military use. The values chosen for each of the application areas are 2, 4, 6, and 8 in the same order. The simulation was then run 1,000 times for the case of a 2G, 3G, 4G and 5G untrusted wireless network. The 1,000 value represent the amount of texts sent. The total successful and unsuccessful messages by attempts and scenario is then reported. The specific values used for the cellular generation types are presented in Table 5.

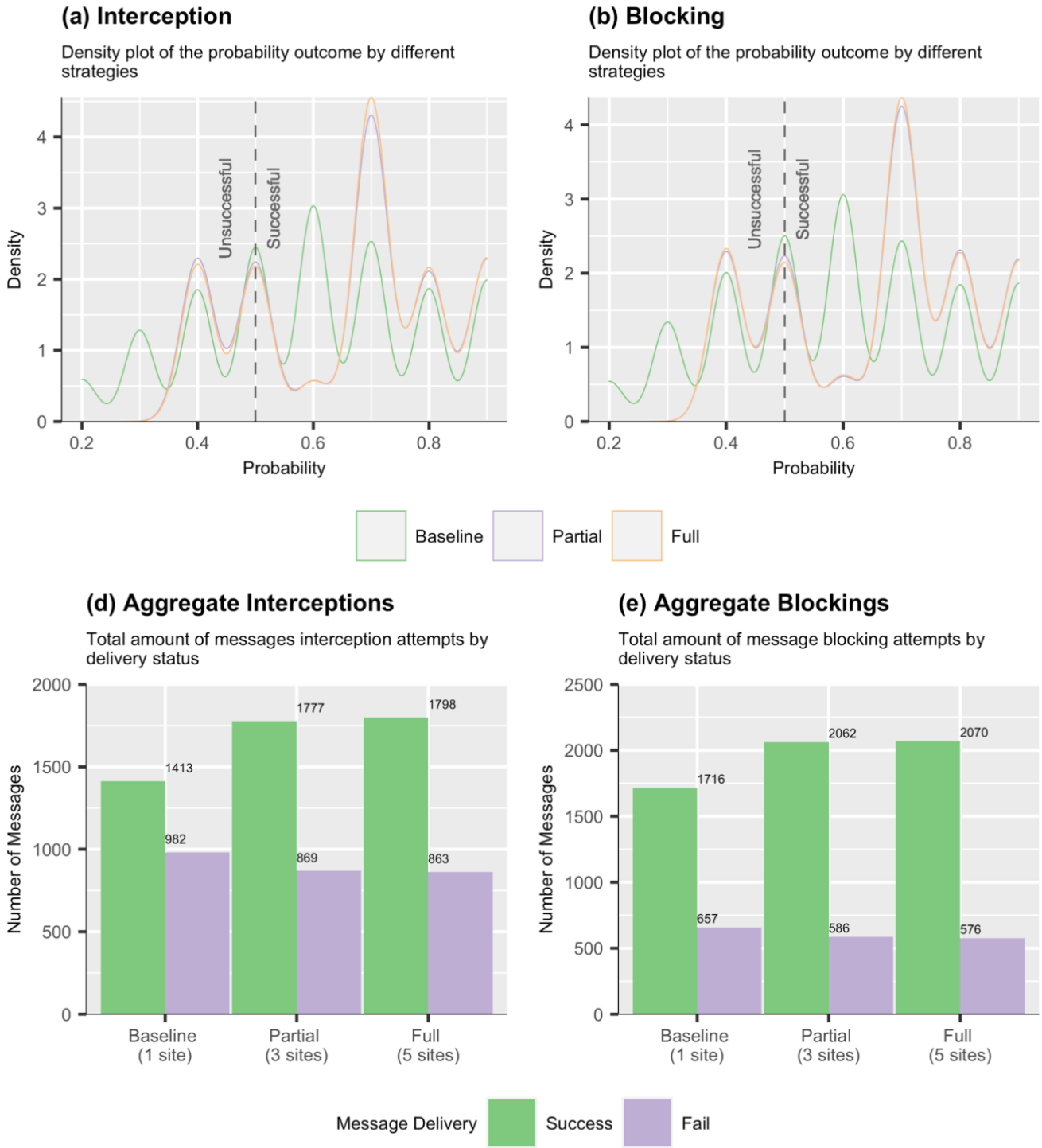

**FIGURE 5.** Simulated probabilistic results of secure texting in a jamming environment.

TABLE 5
USA CELLULAR GENERATION STANDARD DETAILS

| Generation | Frequency | Bandwidth | Transmitted Power |
|---|---|---|---|
| 2G | 850 MHz | 6.8 MHz | 40 dB |
| 3G | 1.9 GHz | 25 MHz | 40 dB |
| 4G | 3.5 GHz | 100 MHz | 40 dB |
| 5G | 26 GHz | 30 GHz | 40 dB |

## IV. RESULTS

In this section, the simulation results described in the application section of the methodology are presented.

### A. JAMMING RESULTS

The jamming general results due to the static interceptor in a 15 $km$ × 15 $km$ grid consisting of 5 untrusted network transmitters are shown in Fig. 4.

The interference power is inversely proportional to the distance between the interceptor and the intended UE (Fig. 4 a). The trend is similar for all the four cellular generation technologies simulated. The case of 2G records the highest partial and 863 for full transmission during interception attempts. For the blocking attempts, the number of unsuccessful messages for baseline, partial and full transmission modes are 657, 586 and 576 respectively.

In terms of probability distribution, the highest density occurs at 0.8. The partial multi-transmission implementation (3 sites) closely follows as most of the probability outcome interference power peaking from 90 dB down to 45 dB compared to 3G (75 dB to 30 dB) and 4G (60 dB to 20 dB). 5G interference has the least power (20 dB to -20 dB).

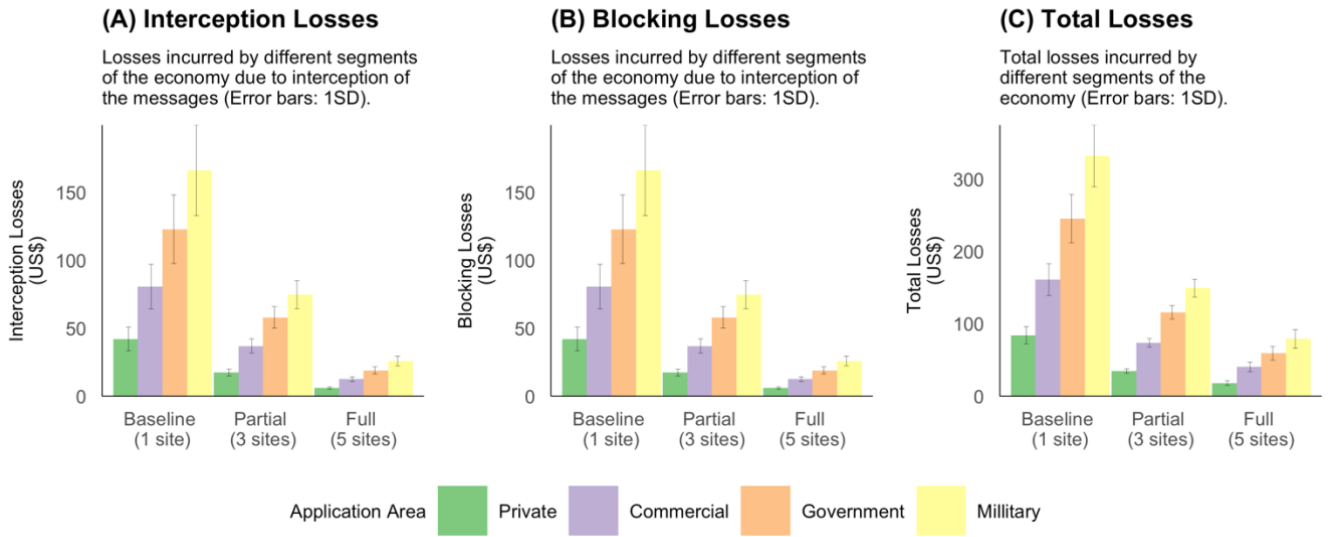

**FIGURE 6.** Simulated cost results due to interception and blocking of mission critical messages by different implementation strategy of anti-jamming techniques.

The power received at the UE is a function of all noise, and unintentional and intentional interference. Fig. 4 b, shows the variation in the power received by different generation technologies. 2G UEs records the highest power range (-35 dB to -57 dB) followed by 3G (-42 dB to -65 dB) and 4G (-47 dB to -70 dB). 5G UEs have the least associated power range (-65 dB to -85 dB).

### B. SECURE TEXTING RESULTS
Fig. 5 represents the density and bar plots of the probability outcomes.

Values above 0.5 indicates failure in intercepting or blocking the messages while the reverse is true for below 0.5. Generally, successful interception or blocking is low for the case where secure texting and full multi-transmission is implemented. A total of 1798 messages are successfully delivered for full compared to 1777 and 1413 of partial and baseline transmissions respectively during interception accept (Fig. 5 c). The same case is evidenced in blocking attempts simulation as shown in Fig. 5 d (full – 2070, partial- 2062, and baseline 1716 messages). Conversely, the number of unsuccessful messages is high for baseline transmission in both interception and blocking attempts compared to partial and full transmission modes. These are 982 for baseline, 869 is concentrated around 0.7 (Fig. 5 a and 5 b). The probability of intercepting/blocking the messages is highest for the case when multi-transmission is not applied (1 site). The peak of the curve at 0.5 and 0.4 indicates the high interception and blocking probability respectively when only one site is used to transmit the message (Fig. 5 a and 5 b).

### C. COST RESULTS
The associated cost results are reported in Fig. 6. The highest interception, blocking, and total losses across all application areas are incurred when the messages are transmitted using only one network. For instance, the total losses for 1,000 texts are US$350 ± 25, US$250 ± 20, US$150 ± 15, and US$75 ± 10 for military, government, commercial, and private users respectively (Fig. 6c). The losses per application area follows a general trend as per the mode of transmission used. Transmission of the messages using all the 5 available sites leads to low interception, blocking and subsequently total losses in all the application areas.

### V. DISCUSSION
The results presented in this paper are part of the open-source framework discussed in the methodology section to model the effect of spatial stochasticity on interference and the corresponding secure texting strategy across selected cellular generations. The approach was used to estimate both interference and receiver power and path loss for a 15 km × 15 km spatial grid scenario with 5 non-cooperative transmitter networks. The results demonstrate the effect of distance on interference in the context of non-cooperative networks. Furthermore, the probabilistic secure texting is modeled to evaluate the effectiveness of such an approach. The results show that transmitting a single message hidden in plain sight over different untrusted non-cooperative networks can be an effective radio security strategy. The probability results formed the basis of simulating the losses associated with intercepting and blocking messages for different application areas. This section now revisits the previously articulated research questions.

### A. WHAT IS THE IMPACT OF ANTI-JAMMING TECHNIQUES ON TEXTING IN NON-COOPERATIVE NETWORKS?

The simulations results agree with the existing jamming theory whereby the interference power is inversely proportional to the interferer-receiver distance. As the interferer-receiver distance shortens, the path loss reduces, resulting to more power from the interfering agent. The high power masks the receiver's meaningful signal resulting to effective signal jamming. Notably, the receiver path loss and subsequent power is a function of the distance to any of the five non-cooperative network transmitters. Therefore, the receivers can limit the level of interference within the grid by transmitting to any close network transmitter and away from the interferer. The least interference occurs at a point where the receiver is as close as possible to the untrusted network and furthest away from the interferer.

In terms of cellular technologies, 2G is more resilient to jamming as opposed to latest generations such as 5G and 4G. This can be attributed to relatively lower frequencies and smaller bandwidths used in this communication protocols. At lower frequencies, the path loss is lower and the propagation distance margin is high. Additionally, the small detection bandwidth reduces the receiver noise resulting in less signal attenuating noise. The general effect of this is higher SINR compared to later technologies (4G/5G).

### B. WHAT IS THE IMPACT OF SECURE TEXTING AND MULTI-NETWORK SENDING IN NON-COOPERATIVE NETWORKS?

The probabilistic success rate of secure texting and multi-network sending in non-cooperative networks as a radio security strategy conforms to existing proposals. High interception and blocking probability outcome occur in scenario where the technique is not applied. However, if the message is hidden in plain sight and transmitted through five different sites, the interception/blocking probability reduces significantly leading to high success rate. Indeed, this is already used in critical applications, such military interventions where adversaries are using multiple civilian networks to relay non-sensitive information. The Russia-Ukraine war has illustrated the usage of untrusted non-cooperative networks to gather intelligence and operate drones [60]. Transmitting a single message through different sites lowers the interception/blocking probability as the risk is spread among each of the cell towers.

The overall results indicate a high message success probability for the application as a radio security measure compared to using a single site. Moreover, the usage of non-cooperative networks enables the involved parties to camouflage in the networks without raising eyebrows. That is not the case for conventional strategies such as encryption where the mere unique military frequency and power signature makes the transmitters an easy target for jamming, blocking and subsequent attack.

### C. WHAT ARE THE SOCIO-ECONOMIC IMPACTS OF THESE TECHNIQUES?

The results from Fig. 6 reveals that the implementation of the two strategies lowers the intercepting, blocking, and total costs. Generally, the losses associated with multi-transmission are lower compared to a case where only one network is used for secure texting. The same is true for partial multi-transmission where only 3 networks are used as opposed to one network.

In terms of application areas, the military and government losses are highest due to the nature and sensitivity of intended messages. The differences in losses with private and commercial usage were reflected in the model to account for the criticality in the former application areas.

Indeed, government and military messages may contain sensitive information such as intelligence reports, briefings, battlefield commands and status updates among others. Interception or blocking of such messages can result into large impacts. Therefore, the simulation results underscore the importance of applying the two techniques to mitigate potential disruption.

## VI. CONCLUSION

The reliability of radio networks is crucial in ensuring that this key technology is available to support dependent applications. However, the nature of their operation makes them vulnerable to attacks by malicious agents. Such attacks can lead to losses in millions of dollars. The cascading effect to the economy and society is unimaginable given the overreliance of human activities on them. Consequently, more efforts have been directed in preventing attacks on them. However, the concentration has been on handling the attack rather than ensuring that it does not occur in the first place. More importantly, limited research exists on quantifying the socio-economic losses associated with radio network attacks.

Motivated by these challenges, this paper presents two mitigation strategy models. The first involves multi-transmission of a single message using untrusted non-cooperative networks. The second model presents a case of secure texting where messages are converted to look like everyday words before they are separated into different bits and transmitted over the multiple untrusted non-cooperative networks. The interception and blocking probability results from the two models forms the basis of quantifying the losses across selected application areas including private, commercial, government and military usage.

The results shows that such mitigation strategies can be an effective approach in avoiding attacks and transmitting messages successfully over untrusted non-cooperative networks. However, the simulation in this paper considers a case of transmission sites with random interception probability. That is not the case in most networks. Moreover, the loss quantification is based on distribution values used in power engineering economics that may not be true for radio network domain. Therefore, further research can improve on the certainty of the interception probability of the sites as well as empirically derived cost values as the codebase is provided [61]. The results form a basis of evaluating the application of untrusted non-cooperative networks and secure texting as a radio security strategy as well as quantifying the associated losses.

### ACKNOWLEDGMENT
We gratefully acknowledge the National Science Foundation (NSF) for funding this research under Award Number 2226423 (NSF 22-538). Also, thanks to anonymous reviewers.

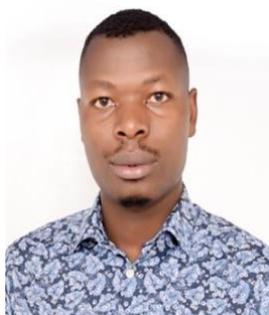

**OGUTU B. OSORO** received the B.Eng. degree in geospatial engineering from the Technical University of Kenya, in 2017, and the M.Sc. degree in satellite applications with data science from the University of Strathclyde, Glasgow, U.K., in 2020. From 2017 to 2018, he was trained by the Development in Africa with Radio Astronomy (DARA) on basics of radio antenna operation, application, signal processing, analysis, and reduction of data. His research interests include signal transmission to LEO, particularly for 5G, and the application of data science and artificial intelligence techniques in satellite.

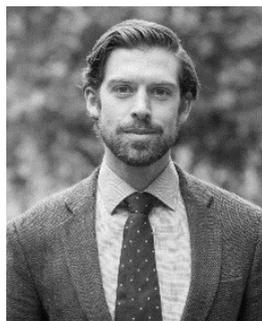

**EDWARD J. OUGHTON** Received the M.Phil. and Ph.D. degrees from Clare College, at the University of Cambridge, U.K., in 2010 and 2015, respectively. He later held research positions at both Cambridge and Oxford. He is currently an Assistant Professor in the College of Science at George Mason University, Fairfax, VA, USA, developing open-source research software to analyze digital infrastructure deployment strategies. He received the Pacific Telecommunication Council Young Scholars Award in 2019, Best Paper Award 2019 from the Society of Risk Analysis, and the TPRC48 Charles Benton Early Career Scholar Award 2021.

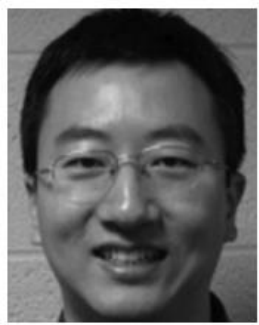

**KAI ZENG**(Member, IEEE) received the Ph.D. degree in electrical and computer engineering from Worcester Polytechnic Institute, Worcester, MA, USA, (WPI), in 2008. He was a Postdoctoral Scholar with the Department of Computer Science, University of California at Davis (UCD), Davis, CA, USA, from 2008 to 2011. He was with the Department of Computer and Information Science, University of Michigan–Dearborn, Dearborn, MI, USA, as an Assistant Professor, from 2011 to 2014. He is currently an Associate Professor with the Department of Electrical and Computer Engineering, Cyber Security Engineering, and the Department of Computer Science, George Mason University, Fairfax, VA, USA. His current research interests include cyber-physical system/IoT security and privacy, 5G and beyond wireless network security, network forensics, machine learning, and spectrum sharing. Dr. Zeng was a recipient of the U.S. National Science Foundation Faculty Early Career Development (CAREER) Award, in 2012, the Excellence in Postdoctoral Research Award from UCD, in 2011, and the Sigma Xi Outstanding Ph.D. Dissertation Award from WPI, in 2008. He is an Editor of the IEEE TRANSACTIONS ON INFORMATION FORENSICS AND SECURITY and IEEE TRANSACTIONS ON COGNITIVE COMMUNICATIONS AND NETWORKING.

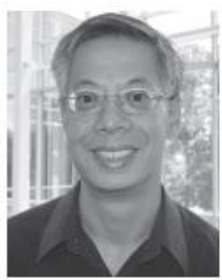

**BRIAN L. MARK**(Senior Member, IEEE) received the B.A.Sc. degree in computer engineering with a minor in mathematics from the University of Waterloo in 1991 and the Ph.D. degree in electrical engineering from Princeton University in 1995. From 1995 to 1999, he was a Research Staff Member with NEC USA, Princeton, NJ, USA. In 1999, he was a Visiting Researcher with Télécom ParisTech, Paris, France. In 2000, he joined George Mason University, where he is currently a Professor of electrical and computer engineering. He has served as the Acting Chair of the Department of Bioengineering from 2015–2017. His main research interests lie in the design and performance analysis of communication networks, wireless communications, statistical signal processing, and network security. He was an Associate Editor of the IEEE TRANSACTIONS ON VEHICULAR TECHNOLOGY from 2006 to 2009.